\documentclass[useAMS,usenatbib,usegraphicx]{mn2e}
\usepackage{amssymb,amsmath,amsfonts,graphicx}
\usepackage{epsfig,color} 

\newcommand{\apjl}{ApJL}
\newcommand{\apjs}{ApJS}

\newcommand{\aap}{A\&A}

\usepackage{times}
\usepackage[bookmarks,bookmarksnumbered,colorlinks=true, citecolor=blue, linkcolor=black]{hyperref}

\title[Satellite Galaxies in Illustris]{The colors of satellite galaxies in the Illustris Simulation}

\author[Sales et al.]{
\parbox[t]{\textwidth}{
Laura V. Sales$^{1}$$\thanks{E-mail: lsales@cfa.harvard.edu}$, 
Mark Vogelsberger$^{2}$, 
Shy Genel$^{1}$,
Paul Torrey$^{1,2}$, 
Dylan Nelson$^{1}$,\\ 
Vicente Rodriguez-Gomez$^{1}$, 
Wenting Wang$^{3}$,
Annalisa Pillepich$^{1}$, 
Debora Sijacki$^{4}$,\\ 
Volker Springel$^{5,6}$ and 
Lars Hernquist$^{1}$
} 
\\
\\
  $^{1}$Harvard-Smithsonian Center for Astrophysics, 60 Garden Street, Cambridge, MA, 02138, USA\\
  $^{2}$Department of Physics, Kavli Institute for Astrophysics and Space Research, Massachusetts Institute of Technology, Cambridge, MA 02139, USA\\
  $^{3}$Institute for Computational Cosmology, University of Durham, South Road, Durham DH1 3LE\\
  $^{4}$Institute of Astronomy and Kavli Institute for Cosmology, University of Cambridge, Madingley Road, Cambridge CB3 0HA, UK\\
  $^{5}$Heidelberg Institute for Theoretical Studies, Schloss-Wolfsbrunnenweg 35, D-69118 Heidelberg, Germany\\
  $^{6}$Zentrum fuer Astronomie der Universitaet Heidelberg, ARI, Moenchhofstr. 12-14, D-69120 Heidelberg, Germany\\
}
\begin{document}



\maketitle

\begin{abstract}

  \noindent Observationally, the fraction of blue satellite galaxies 
  decreases steeply with host halo mass, and their radial distribution
  around central galaxies is significantly shallower in massive
  ($M_* \geq 10^{11} \; \rm M_\odot$) than in Milky Way like
  systems. 
  Theoretical models, based primarily on semi-analytical
  techniques, have had a long-standing problem with reproducing these
  trends, instead predicting too few blue satellites in general but also
  estimating a radial distribution that is too shallow, regardless of
  primary mass. In this Letter, we use the Illustris 
  cosmological simulation to study the properties of
  satellite galaxies around isolated primaries.  For
  the first time, we find good agreement between theory and
  observations.  We identify the main source of this success 
  relative to earlier work to be a consequence of the
  large gas contents of satellites at infall, a factor
  $\sim 5$-$10$ times larger than in semi-analytical
  models.  
  Because of their relatively large gas reservoirs, satellites can
  continue to form stars long after infall, with a typical timescale
  for star-formation to be quenched $\sim 2$ Gyr in groups but more
  than $\sim 5$ Gyr for satellites around Milky Way like primaries.
  The gas contents we infer are consistent with $z=0$ observations of
  HI gas in galaxies, although we find large discrepancies among
  reported values in the literature.  A testable prediction of our
  model is that the gas-to-stellar mass ratio of satellite progenitors
  should vary only weakly with cosmic time.

\end{abstract}


\section{Introduction}
\label{sec:intro}

The observed properties of satellite galaxies can provide insight on a
number of processes related to their environments, and provide
clues about the intrinsic properties of these objects when they were
first accreted by their hosts.
Galaxies that orbit within larger
gravitational potentials are subject to a variety of
physical effects that can significantly decrease their star formation
activity, including tidal and ram-pressure stripping and the
suppression of a supply of fresh gas.
Reddening, mass loss, and morphological transformations
are among the most likely outcomes in response to such
environmental effects.  Moreover, the strength of these processes
will typically decrease at large distance to the host halo center.

\begin{center} \begin{figure*} 
\includegraphics[width=0.49\linewidth]{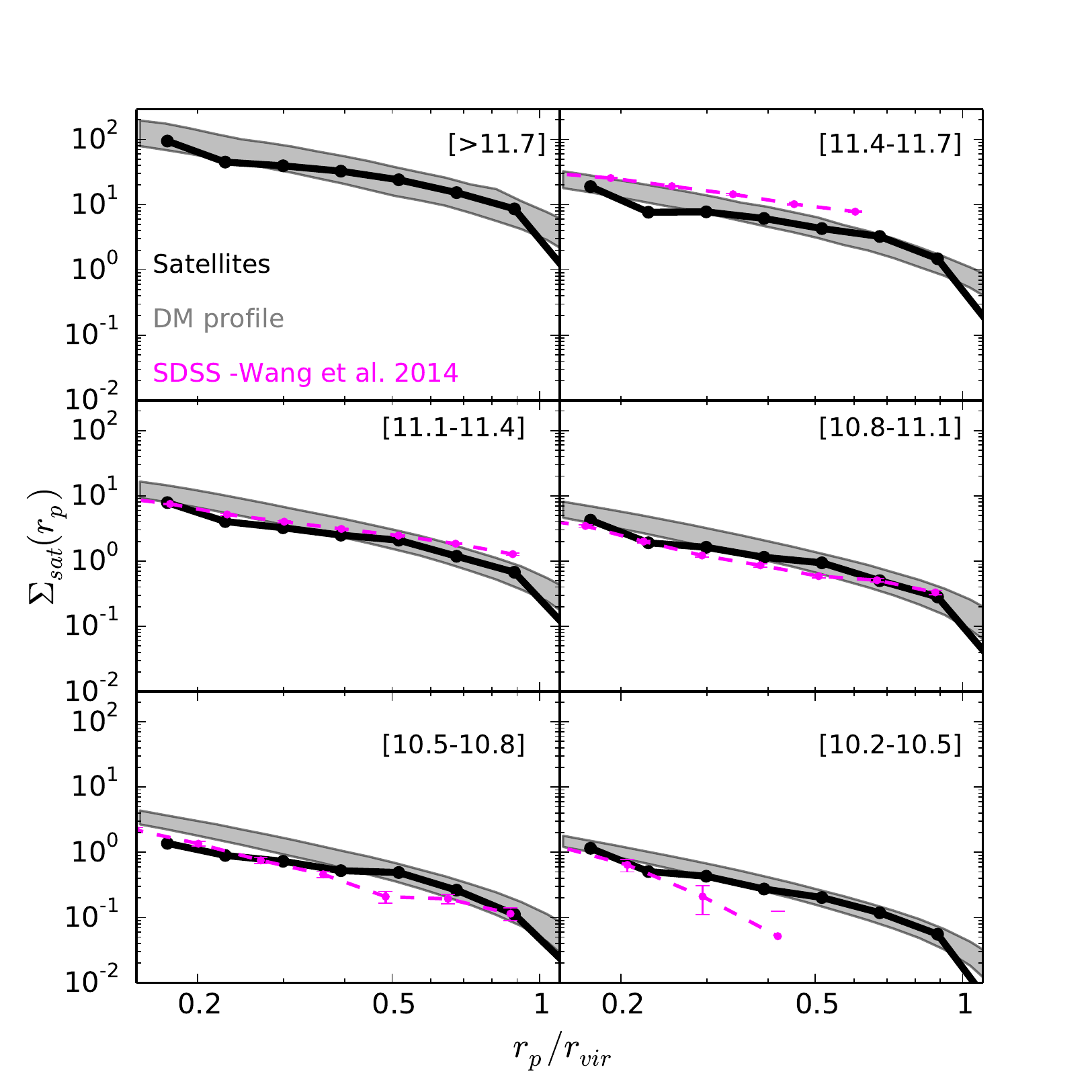}
\includegraphics[width=0.49\linewidth]{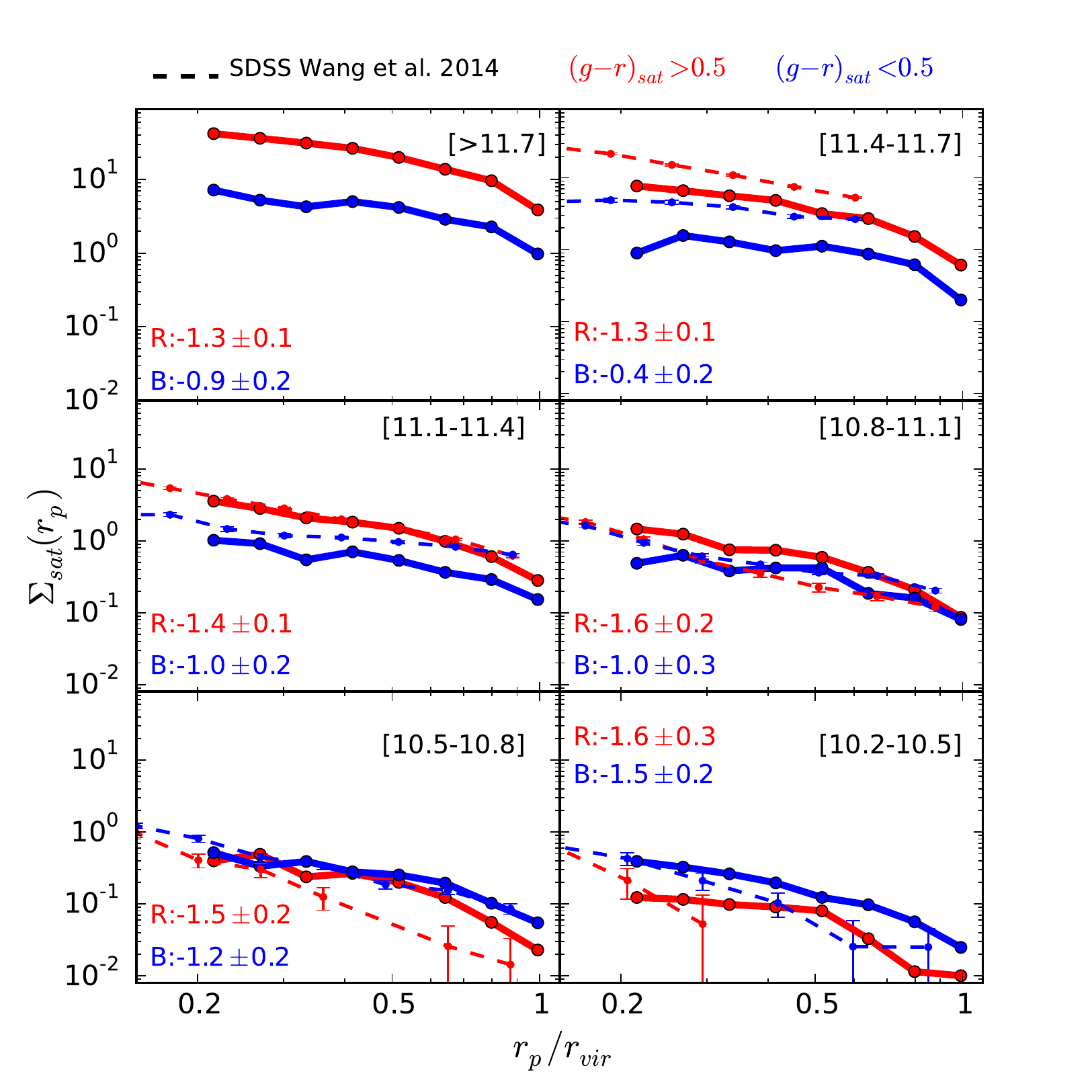}
\caption{{\it Left}: Projected number density profile of
  satellites, $\Sigma_{\rm sat}$,  in Illustris. Panels
  correspond to different primary galaxy stellar masses, as
  indicated (in log-scale). Satellites approximately follow the distribution of
  dark matter particles in their host halos, as shown by the grey
  shaded area corresponding to the 25\%-75\% percentiles of the
  sample (arbitrarily re-normalised such that
  satellites and DM profiles intersect at $r_p/r_{\rm
    vir}=0.5$). Dashed magenta lines show results from the
  SDSS analysis presented in Wang et al. (2014). {\it Right}:
  Same as left panel, but satellites are divided into 
  red and blue populations. Fits to the profiles' slopes are quoted in each panel.
  Our simulations (solid) successfully reproduce two key aspects of the
  SDSS results (dashed): $i)$ a dominant red(blue) population for
  high(low) mass primaries and $ii)$ a steep profile of the blue
  satellites orbiting 
  low mass primaries (bottom row).
}
\label{fig:profiles}
\end{figure*}
\end{center}

Observationally, it is found that
satellite galaxies have a projected number density
distribution close to a power law \citep[e.g.,
][]{Sales2005,vandenBosch2005,Tal2012,Nierenberg2012,Jiang2012},
that is similar to the steep distribution of dark matter
expected in their hosts.  
But, when split by color, it is
expected that blue/star-forming objects should be
under-represented in the inner
regions, leading to a shallower radial distribution of blue satellites
compared to the red population. 
Observations suggest that this is
indeed the case for satellites orbiting groups and clusters with
relatively massive primaries.  However,
for systems with a central galaxy of
intermediate to low stellar mass, most of the satellites 
with stellar masses above $\sim 10^8 \; \rm M_\odot$ are blue,
including those in the inner regions \citep{Prescott2011,Wetzel2012,GuoQuan2013,Wang2014}.

Significant theoretical effort has been devoted to
reproducing these trends with satellite colors. A fair comparison with
observations requires a large number of systems to be analysed,
motivating the use of 
semi-analytical catalogs and {\it ad-hoc}
tagging techniques preferable over hydrodynamic
simulations.
However, these models consistently appear to overestimate
the fraction of red satellites and fail to reproduce the steep
slopes of the blue population around low mass primaries
\citep{Weinmann2006b,Kimm2009,GuoQuan2013,Wang2014}. This has
traditionally been attributed to an overestimate of environmental
effects, possibly related to an improper 
handling of cold/hot gas evolution in satellites
leading to overly rapid quenching
\citep{Font2008,Kang2008,Weinmann2010,Kimm2011}. However,
by suppressing all environmental effects
Wang et al. (2014) recently showed that while this
would increase the
fraction of blue satellites, their radial distribution would still be
significantly shallower than observed.  The main difficulty seems to
be in fueling star formation for several Gyrs after a satellite has
been accreted by its primary host.

In what follows, we examine the distribution of satellites by color 
using the recently completed hydrodynamical
simulation ``Illustris''.  Our results offer a new perspective on
the issues at hand by 
self-consistently following the details of  
internally- and externally-driven evolution of satellites,
explicitly accounting for both their
dark matter and baryonic components.

\section{Numerical Simulations}
\label{sec:simul}

Satellite and primary galaxies are selected from the ``Illustris'' 
hydrodynamical simulation\footnote{http://www.illustris-project.org} 
(Vogelsberger
et al. 2014a,b; Genel et al. 2014).  This simulation is based on a large
cosmic volume ($106.5 \; \rm Mpc$ on a side) with global
properties consistent with a WMAP-9
cosmology \citep{Hinshaw2013} and evolved using the
moving-mesh code
{\sc arepo} \citep{Springel2010}.  Dark matter, gas and stars are
followed from redshift $z=127$ to $z=0$. The simulation includes a treatment of
the astrophysical processes thought to be most important for the
formation and evolution of galaxies, such as gravity, gas
cooling/heating, star formation, mass return and metal enrichment from
stellar evolution, and feedback from stars and supermassive
black holes.  The model reproduces
a number of key observable properties of the galaxy population at the
present-day and at higher redshifts, including stellar mass functions,
scaling relations, color distributions, and the morphological
diversity of galaxies.\nocite{Vogelsberger2014a,Vogelsberger2014b,Torrey2014,Genel2014}

Individual objects in the simulation are identified using the {\sc
  subfind} algorithm \citep{Springel2001,Dolag2009} and are divided
into ``central'' and ``satellite'' galaxies according to their rank
within their friends-of-friends (FoF) group, so that the ``central''
object corresponds (in their majority) to the most massive subhalo of the
group. In what follows, we consider all {\it central} galaxies with
stellar mass $M_*>10^{10} \; \rm M_\odot$ as isolated primaries and
study the distribution and color of their surrounding {\it satellites}
that belong to their FoF groups and that are more massive than $M_*>10^{8}
\; \rm M_\odot$. At the resolution of the simulation (mass per particle
$m_p=6.3 \times 10^6 \;\rm M_\odot$ and $1.3 \times 10^6 \;\rm M_\odot$ for dark matter and baryons,
gravitational softening $\epsilon \sim 500$ at $z=0$),
all such ``satellites'' are resolved with $\sim 100$ or more stellar particles.
Stellar/gas mass and $(g$-$r)$ colors of simulated galaxies are
measured within twice the stellar half mass radius. Virial
quantities correspond to the radius where the spherically-averaged
inner density is $200$ times the critical density of the Universe.

Our simulations track the evolution of satellite galaxies self-consistently,
combining internal processes such as star formation and feedback with external
effects due to the environment,  such as tidal stripping, dynamical friction, ram-pressure, 
gas compression by shocks, etc. The {\sc arepo} code is well suited to handle the
fluid and gravitational instabilities expected to arise in such complex configurations,
with the creation of hot bubbles due to stellar/AGN winds, gaseous tails
due to ram pressure, stellar streams and shells. We have used a lower
resolution box to confirm that the results do not depend strongly on resolution. 
Our sample comprises 9529 satellites around 3306 primary galaxies at $z=0$.

\section{Results}
\label{sec:results}

The left panel of Fig.~\ref{fig:profiles} shows the projected radial
distribution of satellites around primaries in different stellar mass
bins, as indicated in each panel (in log-scale). We consider all
satellites that are within the 3D virial radius of their hosts (taking
all satellites in the FoF group changes only slightly the outer
regions).  The projected profiles, $\Sigma_{\rm sat}$ (black solid
lines), correspond to the average number of satellites found per
primary at a given projected separation $r_p$ (direction of projection
chosen randomly) and normalised to the virial
radius of the host halo $r_{\rm vir}$.
An interesting outcome is seen by comparing this with the shaded grey area
which shows the 25\%-75\% distribution for the dark matter
profiles in these halos (normalised arbitrarily so that they
intersect at $r_p/r_{\rm vir}=0.5$). We find that satellite galaxies 
roughly follow the underlying distribution of dark matter in the distance range 
$0.2 \leq r_p/r_{\rm vir}<1$, with a trend to flatten towards
the inner regions. Previous works have shown no consensus 
on this issue, although most of the discrepancies might be explained 
by different selection criteria and distance range considered 
(see Section 1,  Wang et al. (2014)).
%
\begin{center} \begin{figure} 
\includegraphics[width=90mm]{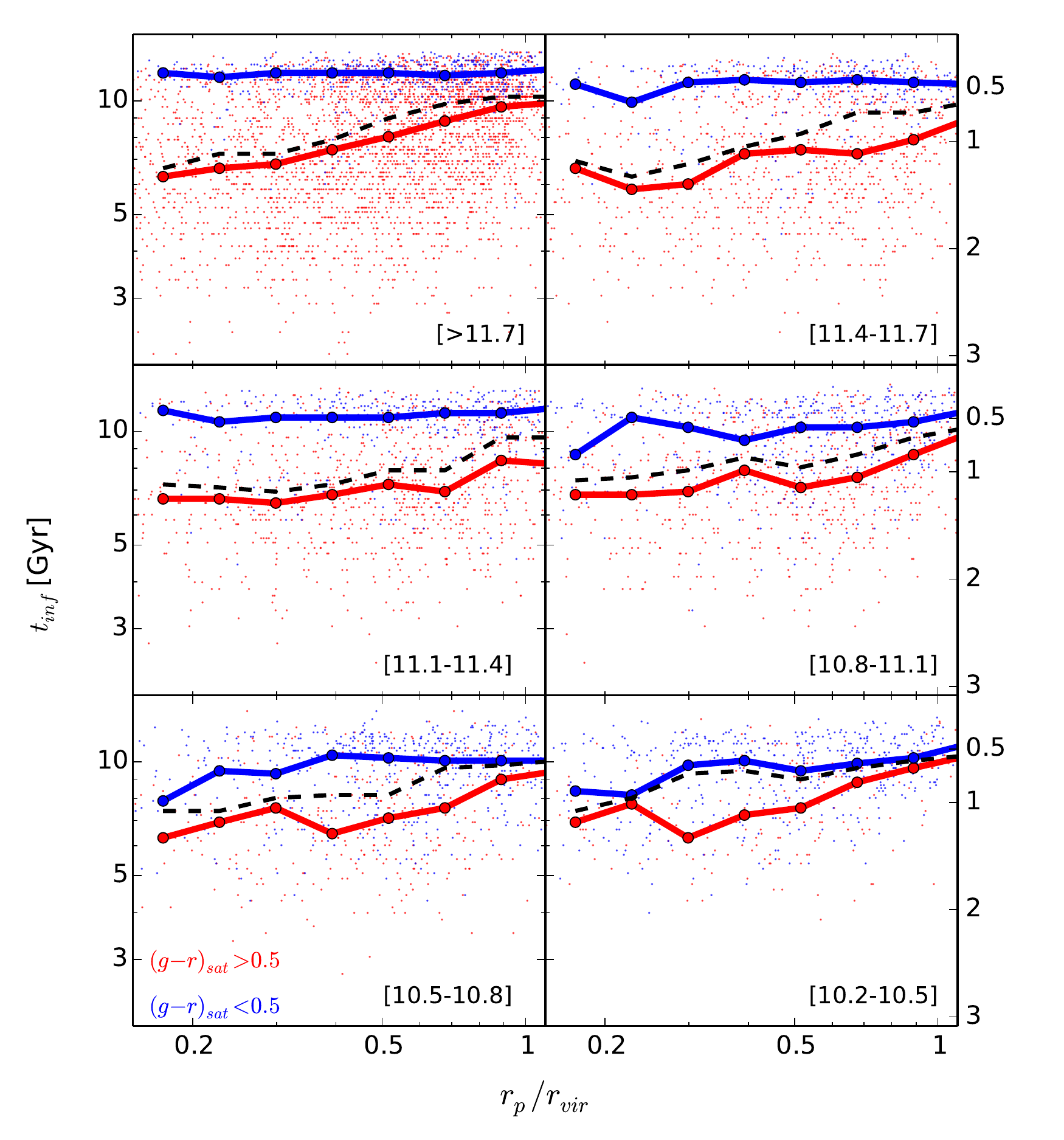}
\caption{Infall times $t_{\rm inf}$ for satellites as a function of
  present-day projected distance (right $y$-axes display corresponding redshifts). 
  Points indicate individual
  satellites and red/blue corresponds to their $(g$-$r)$ colors at
  $z=0$. The median trend considering all satellites is shown by the
  black dashed lines and is roughly independent of the primary stellar
  mass (different panels). However, when split by color (solid curves), satellites
  that remain blue today can have significantly earlier infall times in low mass
  primaries (bottom two panels) than in more massive hosts.}
\label{fig:tinf}
\end{figure}
\end{center}

The left panel of Fig.~\ref{fig:profiles} also shows (dashed magenta 
lines) observational results based on photometric and
spectroscopic SDSS data from Wang et al. (2014). In this work,
primaries are identified following
isolation criteria for their projected
distance and redshift distributions, while satellite profiles result
from the photometric SDSS sample after properly accounting for
background/foreground objects through a subtraction method.  Stellar
mass cuts in primaries and satellites are similar to our 
analysis.  The
good agreement between the slopes and normalisations of the black solid
and dashed magenta lines is encouraging, especially taking into
account the different selection criteria used in the two 
samples. (For the least massive bins, the background subtraction
method and the isolation criteria can have a significant
impact on the profile slopes obtained; see Appendix in Wang et al.)

We explore satellite profiles split by color in the right panel of
Fig.~\ref{fig:profiles}.  Here, we adopt a uniform color cut $(g-r)=0.5$
independent of satellite mass, but we have explicitly checked that
this choice does not qualitatively bias our results.  For the most
massive primaries ($\rm log$$(M_*/\rm M_\odot)>11$) red satellites dominate the
overall population and tend to be distributed more steeply than the blue
ones, especially for the four most massive primary bins. However, in
lower mass systems, satellites are predominantly blue and exhibit a
steeper radial profile than in more massive systems. 
We quote in each panel the slope found for each 
population and its uncertainty based on 100 bootstrap resamplings of the data.
 These trends agree well with observations,
including the SDSS results of Wang et
al. (2014) -- shown here by the red/blue dashed lines (see also Guo et
al. 2013).  Reproducing these behaviours and the abundance of blue
satellites around low mass primaries has been a challenge in theoretical
models of galaxy formation based on
semi-analytical methods, but they seem to arise naturally
in our simulations. 

Variations in satellite infall times 
for a wide range of 
primary masses could explain their different color distributions.
However, we find that $t_{\rm inf}$ is roughly independent of primary
mass, as shown by the median relations (black dashed curves) in
Fig.~\ref{fig:tinf}.  We define $t_{\rm inf}$ as the last time a
satellite was a central galaxy of its own FoF group, and explicitly
checked that using a different definition (e.g., the time when the
satellite joins the FoF group to which it belongs at $z=0$) yields
similar results. When we split satellites by their own colors we find
large differences. In the most massive systems, regardless of the
projected distance of the satellite, blue objects have typically 
fallen in only
recently, fewer than $\sim 2$ Gyrs ago. Instead, for primaries
comparable to the Milky Way, the median infall time of blue satellites
in the inner regions is $\sim 5$ Gyrs ago, with individual cases
scattering down to $>7$ Gyr.
%
\begin{center} \begin{figure} 
\includegraphics[width=90mm]{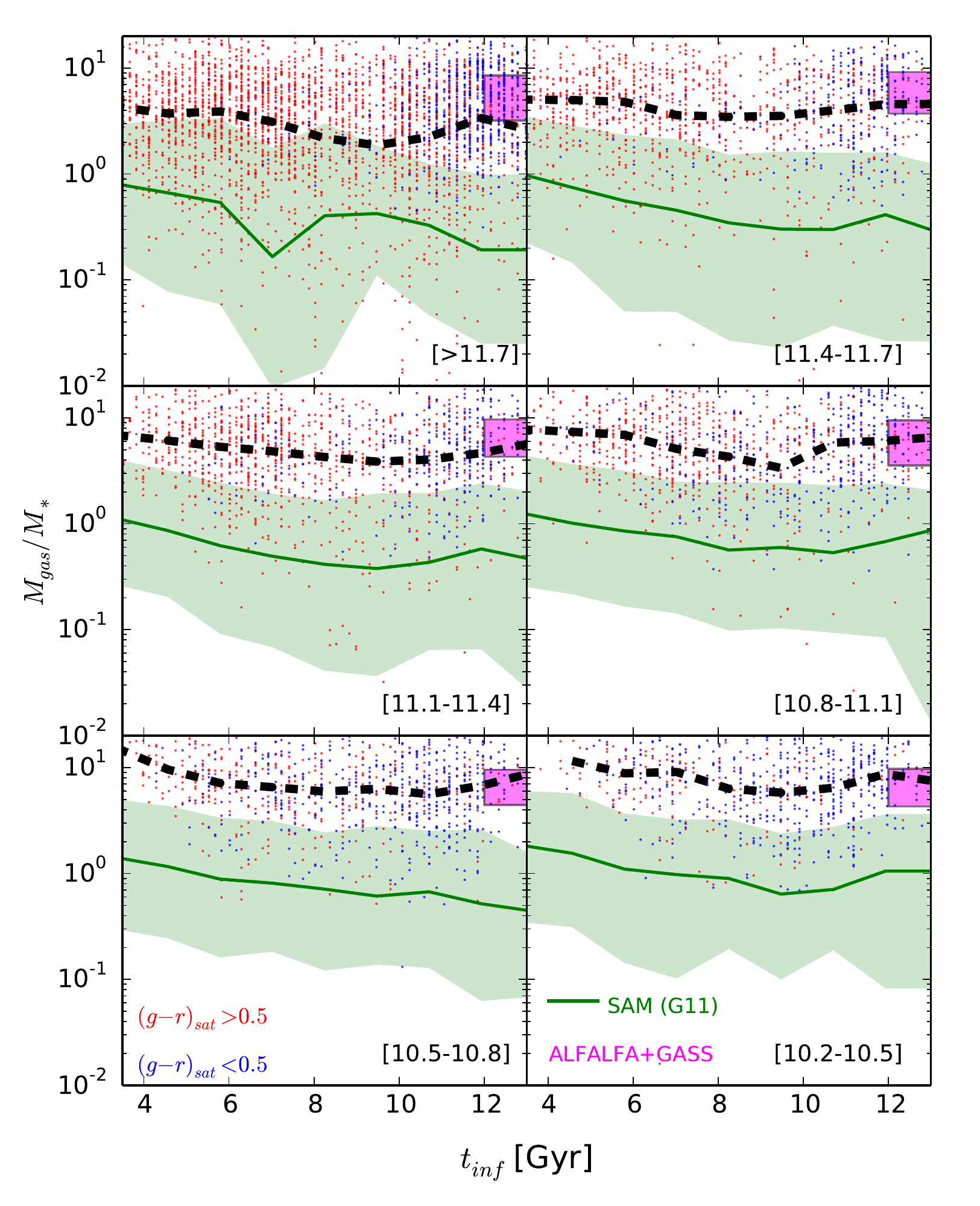}
\caption{Satellite gas-to-stellar mass ratios
  at infall, $M_{\rm gas}/M_*$, as a function of infall time. As earlier, the
  panels correspond to different primary stellar masses, red/blue denotes
  $(g$-$r)$ colors of satellites at $z=0$ and the median trend is
  shown by the black dashed curves. 
  (The vertical stripes in the points distribution reflect the finite number of 
  output times.) In Illustris, satellites infall with a
  large gas content ($M_{\rm gas}/M_* \sim2$-$8$) that allows them to
  form stars for several Gyrs. The gas contents in our model are, at
  present epoch, in good agreement with observational estimates from the
  {\sc ALFALFA} and {\sc GASS} surveys (magenta rectangles; see text for more
  details). The green shaded region indicates gas-to-stellar mass
  ratios adopted in
  semi-analytical models (in this case taken from Guo et
  al. 2011), which are substantially lower than in our
  simulations. This appears to be a key factor to explaining the
  overproduction of red satellites in SAMs.}
\label{fig:fgas}
\end{figure}
\end{center}

Interestingly, red satellites cease forming stars on very 
different timescales according to their orbits: they show slow gas 
consumption in wide orbits to rapid single pericenter episodes 
for very radial ones. In general, surviving satellites
maintain a similar stellar mass to that at infall, although
a few percent ($2\%$-$10\%$, depending on primary mass) more 
than double their stellar content during their lives as satellites
(see also Pillepich et al. 2014). \nocite{Pillepich2014a} 
Appreciable stellar stripping occurs only for red satellites, but 
the impact is small; fewer than $3\%$ have lost at least half their 
initial stellar mass. These results are in reasonably good agreement 
with the standard treatment of star formation in satellite galaxies 
employed in semi-analytical models.

We find that a key factor in achieving a prolongued star formation
for satellites in Fig. 2 (and therefore, producing the good match 
between observed satellite
colors/distribution) is the large gas contents of satellites at
infall. Fig.~\ref{fig:fgas} shows, in bins of primary stellar mass,
the infall gas-to-stellar mass ratios $M_{\rm gas}/M_*$ of satellites
as a function of their infall times. We consider all gas particles within
twice the stellar half-mass radius and we have checked that
it is dominated (in more than 90$\%$) by the cold ($T<2\times 10^4 \; ^\circ\rm K$)
plus star-forming components. The median in our simulations
suggests that, at infall, satellites have $\sim 2$-$5$ times more 
gas than stars.
We also find a weak trend of $M_{\rm gas}/M_*$ with the
primary stellar mass (different panels), that is a consequence of the
most massive systems having more massive satellites, which are, on
average, less gas-rich than the lower mass dwarfs that dominate the
satellite population of intermediate and low mass primaries (see Fig.3
in Vogelsberger et al. (2014a) for the mass-dependent nature of gas
contents in our simulations and observations).

To relate our finding with previous results in the literature, we
compare our simulated gas contents with those from the publicly
available semi-analytical model of \citet{Guo2011}, which has itself
(or its predecessors) been extensively used in the study of satellite
galaxy properties. We select primaries and satellites from the
Millennium-II volume \citep{Boylan-Kolchin2009} with the exact same
selection criteria as employed in analysing our simulations. The
median (green solid line) and $10\%$-$90\%$ percentiles (shaded area)
of the distribution of cold gas-to-stellar mass ratios in the SAM are
shown in Fig.~\ref{fig:fgas}. The difference with the simulated data
is striking, in the sense that the galaxies are approximately a factor
$10$ times less gas-rich in the semi-analytical catalog than in
Illustris. We have checked that this is independent of satellite
stellar mass and is also present for {\it central} galaxies in
  the model.  On this basis, we argue that the relatively low gas
content of satellites at infall in the semi-analytical models is
largely responsible for their failure to reproduce the color and
distribution of satellite galaxies.

Unbiased observational estimates of gas contents are challenging,
specially for low mass objects. Gas fractions reported in the
literature show large variations depending on the observational
strategy. Optically selected samples find systematically lower gas
fractions than HI-selected samples like {\sc ALFALFA} \citep[see
][]{Papastergis2012}. Ideally, the best constraints should come from
mass-complete, optically selected samples like {\sc GASS}
\citep{Catinella2010}.  However, presently these surveys are restricted to
only massive galaxies
($M_*>10^{10} \; \rm M_\odot$). Because low-mass central galaxies are
mostly star-forming, we assume that satellites before infall follow
{\sc ALFALFA} reported gas fractions from \citet{Huang2012} if
$M_*<10^{10} \; \rm M_\odot$ but that of {\sc GASS} if they are more
massive. The magenta rectangles in Fig.~\ref{fig:fgas} indicate the
gas-to-stellar mass ratios obtained by assigning to each simulated
satellite with a recent infall time ($t_{\rm inf}>12 \; \rm Gyr$) the
gas content according to this {\sc ALFALFA}+{\sc GASS} combination, a
$0.3$ dex dispersion around the mean, and assuming a mass-independent
molecular fraction of 0.3 \citep{Saintonge2011}. The simulated values
are in good agreement with what would be expected from such
observations at $z=0$.

Lastly,  since the parameters in the
model used in Illustris were not calibrated to reproduce gas quantities,
the weak dependence of the (total) gas-to-stellar ratio with infall redshift is actually
a prediction of our model. Interestingly, an independent analysis of
the stellar mass - metallicity relations at several redshifts
\citep{Zahid2014} also seems to indicate such a weak evolution.

\section{Conclusion}
\label{sec:concl}

We use the Illustris cosmological simulation to study
the distribution of satellite galaxies around isolated primaries. The
large volume of our simulated box allows us to statistically characterise
$\sim 9500$ satellite galaxies with $M_*>10^8 \; \rm M_\odot$
orbiting primaries with masses comparable to the Milky Way and above.
This is the first time such a study has been performed using a
hydrodynamical simulation. 

We find good agreement between our results and observations from
the SDSS wide field survey.  In our simulations,
$i)$ satellites roughly trace the distribution of dark matter
in their hosts, and $ii)$ in high-mass systems, red satellites dominate 
and are distributed more steeply than the blue population, whereas for lower
mass primaries, satellites are mostly blue and they also follow a
steep number density profile. This good agreement with the observations
contrasts with earlier theoretical modeling, mostly based on
semi-analytical techniques, which were unable to
reproduce the satellite color -
primary mass dependent behaviour seen in the observations. 
Based on the semi-analytical model output of Guo et al. (2011), we suggest that
gas contents that are too low for satellites at infall (and not
the modelling of environmental effects) is
the most likely cause of the challenges faced by such models in
reproducing observations.

Our simulations provide a tool for understanding the timescales for
the quenching of star formation in different environments. At infall,
simulated
satellites carry significant amounts of gas, with quartiles $M_{\rm gas}/M_*=2$-$8$, 
that can fuel star formation for long periods of time.
Moreover, we find a very weak evolution
of gas-to-stellar mass ratios with redshift, a testable prediction of the model
that can be explored once observational estimates of total gas mass
become available for galaxies at higher redshifts. 

%
%
\section*{Acknowledgements}
We are grateful to the anonymous referee for a constructive and positive
report that helped improve the previous version of this letter. 
LVS wishes to thank Simon White and Guinnevere Kauffmann for
stimulating discussions and Manolis Papastergis for facilitating 
ALFALFA tables. Simulations were run on the Harvard
Odyssey and CfA/ITC clusters, the Ranger and Stampede supercomputers
at the Texas Advanced Computing Center as part of
XSEDE, the Kraken supercomputer at Oak Ridge National Laboratory
as part of XSEDE, the CURIE supercomputer at CEA/France
as part of PRACE project RA0844, and the SuperMUC computer at
the Leibniz Computing Centre, Germany, as part of project pr85je.
L.H. acknowledges support from NASA grant NNX12AC67G and NSF
grant AST-1312095. V.S. acknowledges support from the European
Research Council through ERC-StG grant EXAGAL-308037.


\bibliography{master}

\end{document}